\documentclass{article}
\usepackage{spconf,amsmath,graphicx}
\usepackage{xcolor}
\usepackage{amssymb}
\usepackage{url}
\usepackage{hyperref}

\input listofitems
\let\cpar\relax
\def\Centerline#1{%
  \setsepchar{\cpar}%
  \readlist\clarg{#1}%
  \foreachitem\z\in\clarg[]{\centerline{\z}}%
}
\title{EEG2IMAGE: Image Reconstruction from EEG Brain Signals}

\name{Prajwal Singh\thanks{This work is supported by Prime Minister Research Fellowship (PMRF-2122-2557) to PS and thanks to SERB and PlayPowerLabs for PMRF to PP. We thank FICCI for facilitating the PMRF of PP.}$^{\star}$ \qquad Pankaj Pandey$^{\dagger}$ \qquad Krishna Miyapuram$^{\dagger}$ \qquad Shanmuganathan Raman$^{\star}$}

			\address{$^{\star}$ CVIG Lab, $^{\dagger}$ Brain Lab \\
			    $^{\star \dagger}$Indian Institute of Technology Gandhinagar, India \\ \{ singh\_prajwal, pankaj.p, kprasad,
                shanmuga\}@iitgn.ac.in}

\begin{document}
\ninept
\maketitle

\begin{abstract}
\label{sec:abstract}

Reconstructing images using brain signals of imagined visuals may provide an augmented vision to the disabled, leading to the advancement of Brain-Computer Interface (BCI) technology. The recent progress in deep learning has boosted the study area of synthesizing images from brain signals using Generative Adversarial Networks (GAN). In this work, we have proposed a framework for synthesizing the images from the brain activity recorded by an electroencephalogram (EEG) using small-size EEG datasets. This brain activity is recorded from the subject's head scalp using EEG when they ask to visualize certain classes of Objects and English characters. We use a contrastive learning method in the proposed framework to extract features from EEG signals and synthesize the images from extracted features using conditional GAN. We modify the loss function to train the GAN, which enables it to synthesize $128 \times 128$ images using a small number of images. Further, we conduct ablation studies and experiments to show the effectiveness of our proposed framework over other state-of-the-art methods using the small EEG dataset.
\end{abstract}

\begin{keywords}
Deep Learning, EEG, GAN
\end{keywords}
\section{Introduction}
\label{sec:intro}
Human visual system is considered a highly advanced intelligent information processor that generates rich 3D visuals with semantic construction. The most challenging problem is to train artificial machines to construct images from brain activity directly to semantic categories \cite{miyawaki2008visual}. The possibility of brain-to-image construction significantly contributes to advancing the Brain-Computer Interface (BCI) technology. The core purpose of BCI using invasive or non-invasive techniques is to provide communication and control of external devices by thought alone or using minimal muscular activity. This effort would be highly relevant in neuro-rehabilitation, i.e. to support patients with disabilities to have better everyday communication in their lives. Decoding brain responses to imagination/visual stimuli would greatly benefit the communication exchange for the disabled people. The most widely employed brain imaging modality with high temporal precision is Electroencephalography (EEG) due to its relatively lower cost and portability.

EEG is a non-invasive technique which makes it the most practical methodology to record the electrophysiological dynamics of the brain. EEG signals have been used to analyze a wide spectrum of research from aspects of cognition to clinical aspects \cite{sakkalis2011applied}. For decades, EEG signals have been widely employed for classifying several disorders or understanding brain dynamics. The successful implication of the past results can be seen in the BCI. A previous effort made by the research community has shown promising results to augment healthy individuals with additional sensory or motor capabilities \cite{bamdad2015application}. The most intriguing task is to decode the content of the mind using brain signals and draw a link between them. Two most challenging endeavors in this space are to reconstruct the visualized images \cite{kavasidis2017brain2image} and decode imagined speech-to-text \cite{wang2022open}  based solely on recorded brain signals.
Vision neuroscientists made the initial attempts \cite {carlson2013representational,carlson2011high,das2010predicting} to provide an evidence of visual stimuli features represented in recorded brain activity. These attempts initiated the classification of image categories using brain signals using deep learning and further led to the reconstruction and generation of the images \cite{spampinato2017deep}. 

\begin{figure}[!t]
\begin{minipage}[b]{1.0\linewidth}
  \centering
  \centerline{\includegraphics[width=7.5cm]{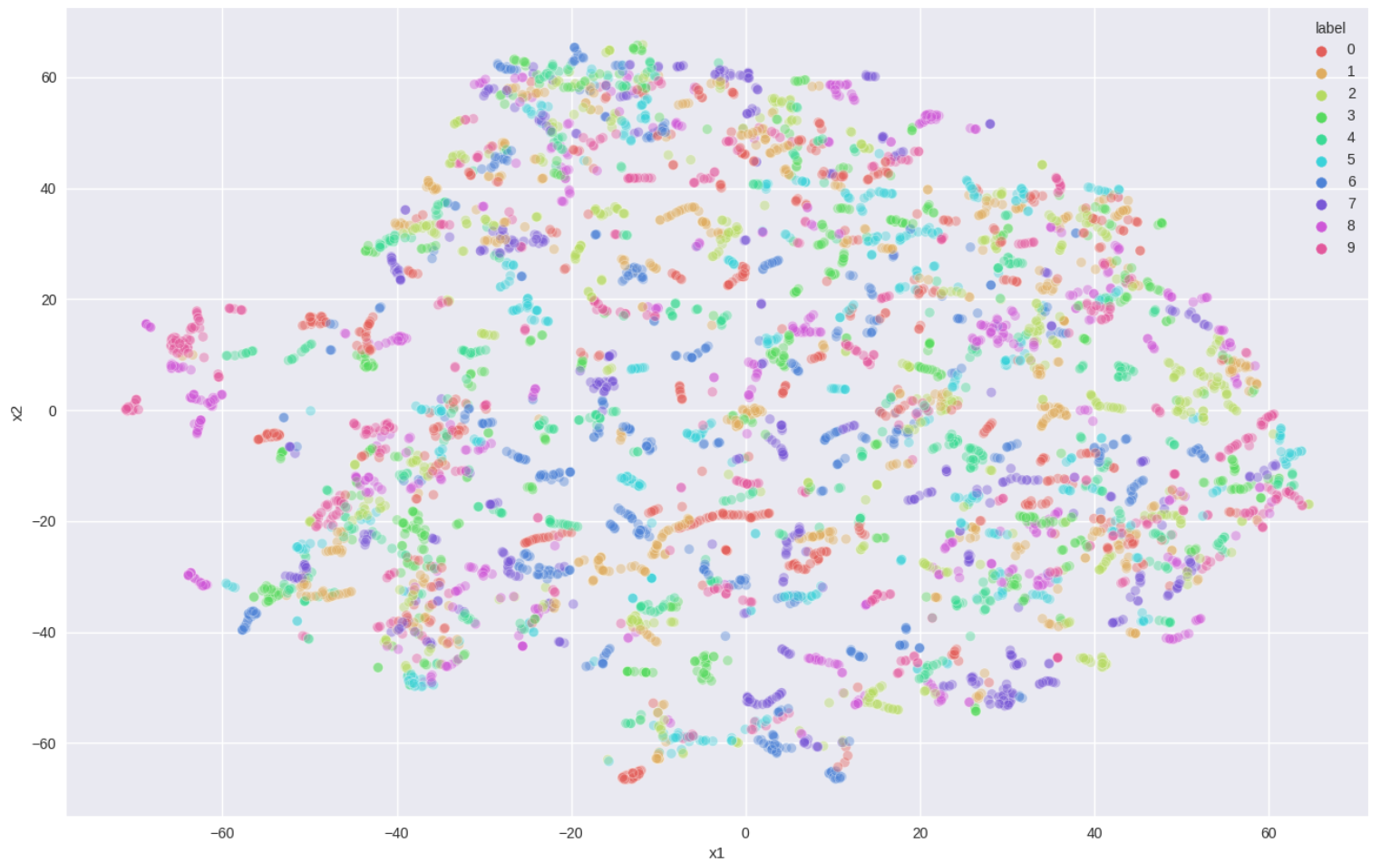}}
\end{minipage}
\caption{t-SNE \cite{van2008visualizing} visualization of Object test dataset \cite{Kumar2018} EEG feature space which is learned using label supervision with test classification accuracy 0.75 and k-means accuracy 0.18.}
\label{fig:class_feature}
\vspace{-3.5mm}
\end{figure}

Our contributions are as follows: 1) A framework that can synthesize images using a small EEG dataset, 2) Use of semi-hard triplet loss \cite{schroff2015facenettriplet} to learn features from EEG signals that show better k-means accuracy than the softmax counterpart, as shown in Figs. [\ref{fig:class_feature},\ref{fig:res_feature}] and 3) Use of mode seeking regularization \cite{mao2019mode} and data augmentation \cite{diffaug} based modification to GAN for synthesizing high-quality images using conditional GAN as shown in Fig. \ref{fig:res_imagenet10}(b).\footnote{\url{https://github.com/prajwalsingh/EEG2Image}}

\begin{figure*}[!t]

\begin{minipage}[b]{0.3\linewidth}
  \centering
  \centerline{\includegraphics[width=5.0cm]{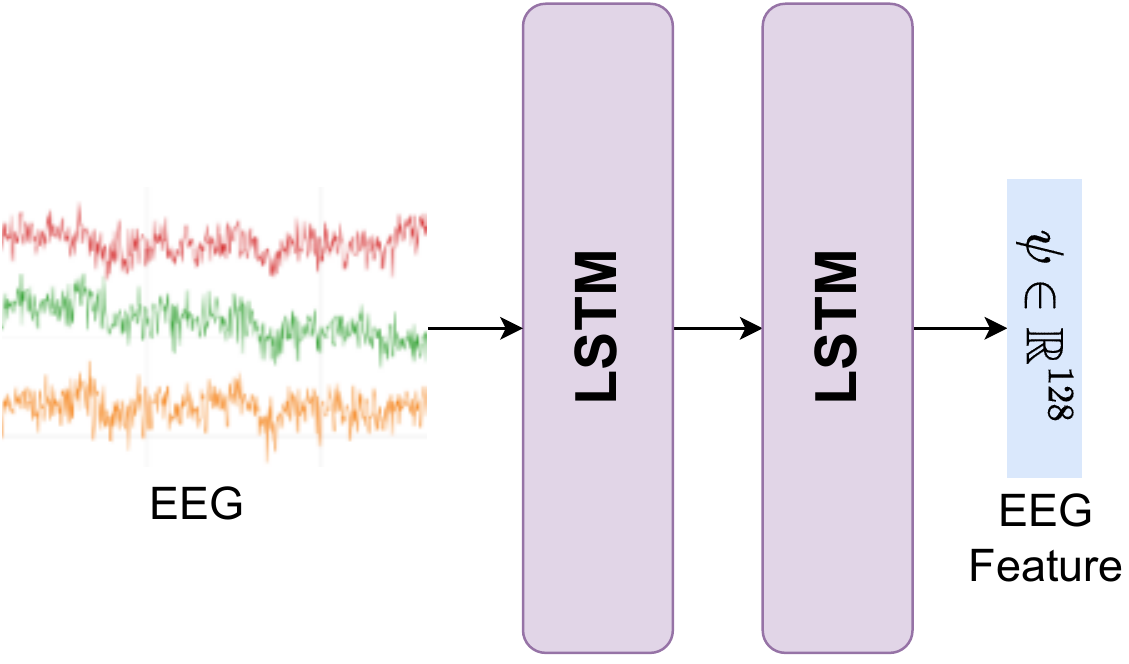}}
  \centerline{(a) EEG feature extractor}\medskip
\vspace{-3mm}
\end{minipage}
%
\begin{minipage}[b]{0.7\linewidth}
  \centering
  \centerline{\includegraphics[width=10.5cm]{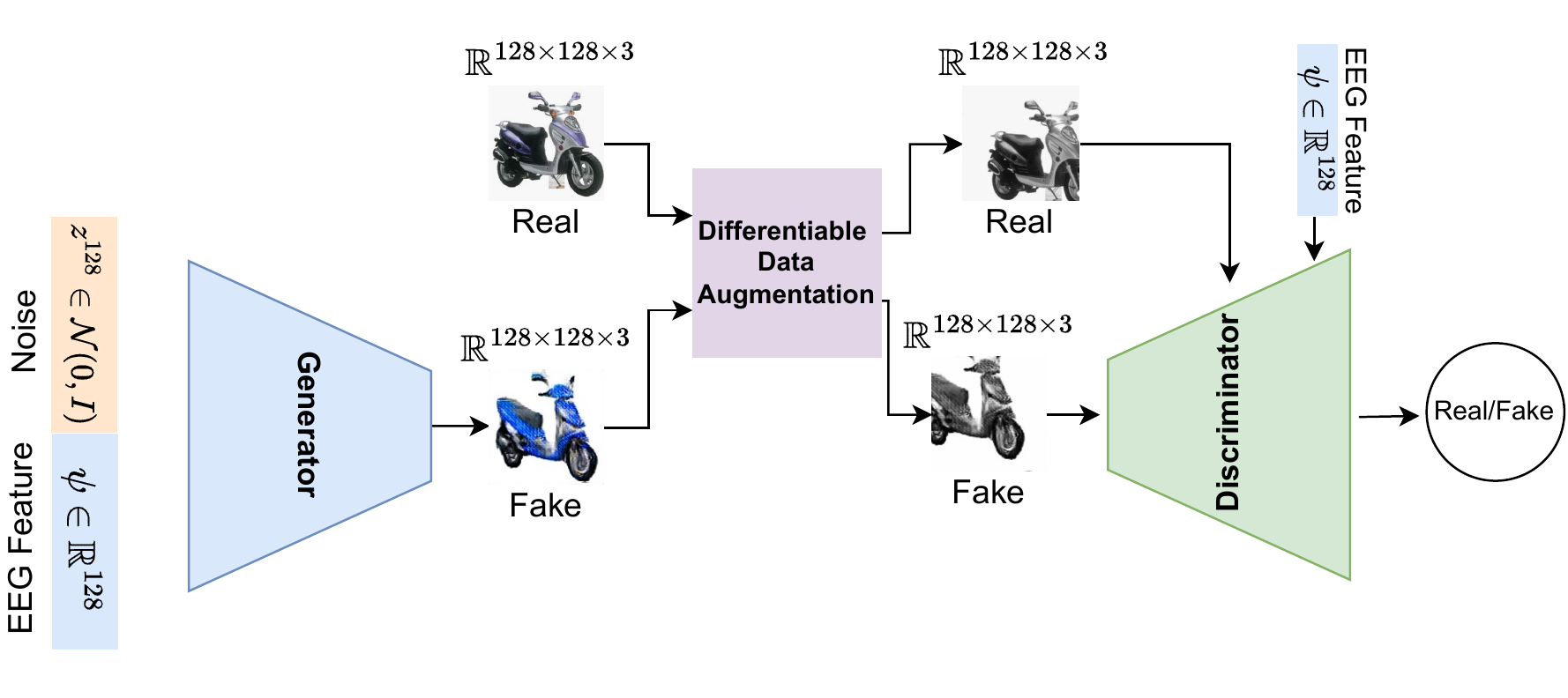}}
  \centerline{(b) EEG2Image}\medskip
\vspace{-3mm}
\end{minipage}
\caption{This figure illustrates the proposed framework for EEG feature extraction and image generation. a) Shows the LSTM network with $128$ hidden units that transforms EEG signal into $128D$ feature vector. b) Shows the GAN network with a data augmentation block that prevents the discriminator from memorizing the small dataset and helps the generator synthesize high-quality images.}
\label{fig:network}
\vspace{-3mm}
\end{figure*}
\vspace{-0.1cm}
\section{Related Works}
\label{sec:relatedworks}
The development of advanced deep generative architectures in recent times has made it possible to see images from brain signals. The initial study by Kavasidis \emph{et al.} implemented long short-term memory (LSTM) stacked with generative techniques to generate seen images from 40 Image Net classes \cite{kavasidis2017brain2image}. Thoughtviz \cite{thoughtviz} encouraged the design of conditional GAN (cGAN) to decode EEG signals using a small dataset consisting of imagination tasks comprising digits, characters, and objects. Several architectures have been developed using CNN and LSTM on the time-series data of most biological areas. The capability of LSTM for identifying the sequential pattern and CNN to locate the neighborhood features was recently combined with spectral normalization generative adversarial network (SNGAN) to yield seen images from EEG encodings \cite{zheng2020decoding}. Researchers are putting effort into reconstructing geometrical shapes from brain activities, primarily in generating precise edges and other low-level details. Further advancement in GAN leads to synthesizing natural geometrical shapes, which enforces semantic alignment constraints to construct natural shapes at pixel-level \cite{zhang2019multi,fares2020brain}. Recently, a siamese network was utilized to maximize the relationship between extracted manifold brain feature representation and visual features \cite{palazzo2020decoding}. The obtained representation demonstrated better image classification and saliency detection performance on the learned manifold. Khare \emph{et al.} proposed conditional progressive growing of GANs (CProGAN) to develop perceived images \cite{khare2022neurovision} and showed higher inception than previous related work. Recent work on contrastive self-supervised approach has been shown to maximize the mutual information between visual stimulus and corresponding EEG latent representations \cite{ye2022see}. They proposed an approach that employed cross-modal alignment enforcing image retrieval at the instance level rather than pixel-level generation.

\vspace{-2mm}
\section{Proposed Method}
\label{sec:method}
In this work, we proposed a framework shown in Fig.\ref{fig:network}, for visualizing the brain activity EEG signals. The framework consists of a two-phase approach: 1) extracting good features from the EEG signals with a contrastive learning approach and 2) a conditional data-efficient GAN to transform the extracted EEG features to image. In our case, a good feature implies useful information about an image that can help GAN to reconstruct that image.
\begin{figure}[!t]
\begin{minipage}[b]{1.0\linewidth}
  \centering
  \centerline{\includegraphics[width=7.5cm]{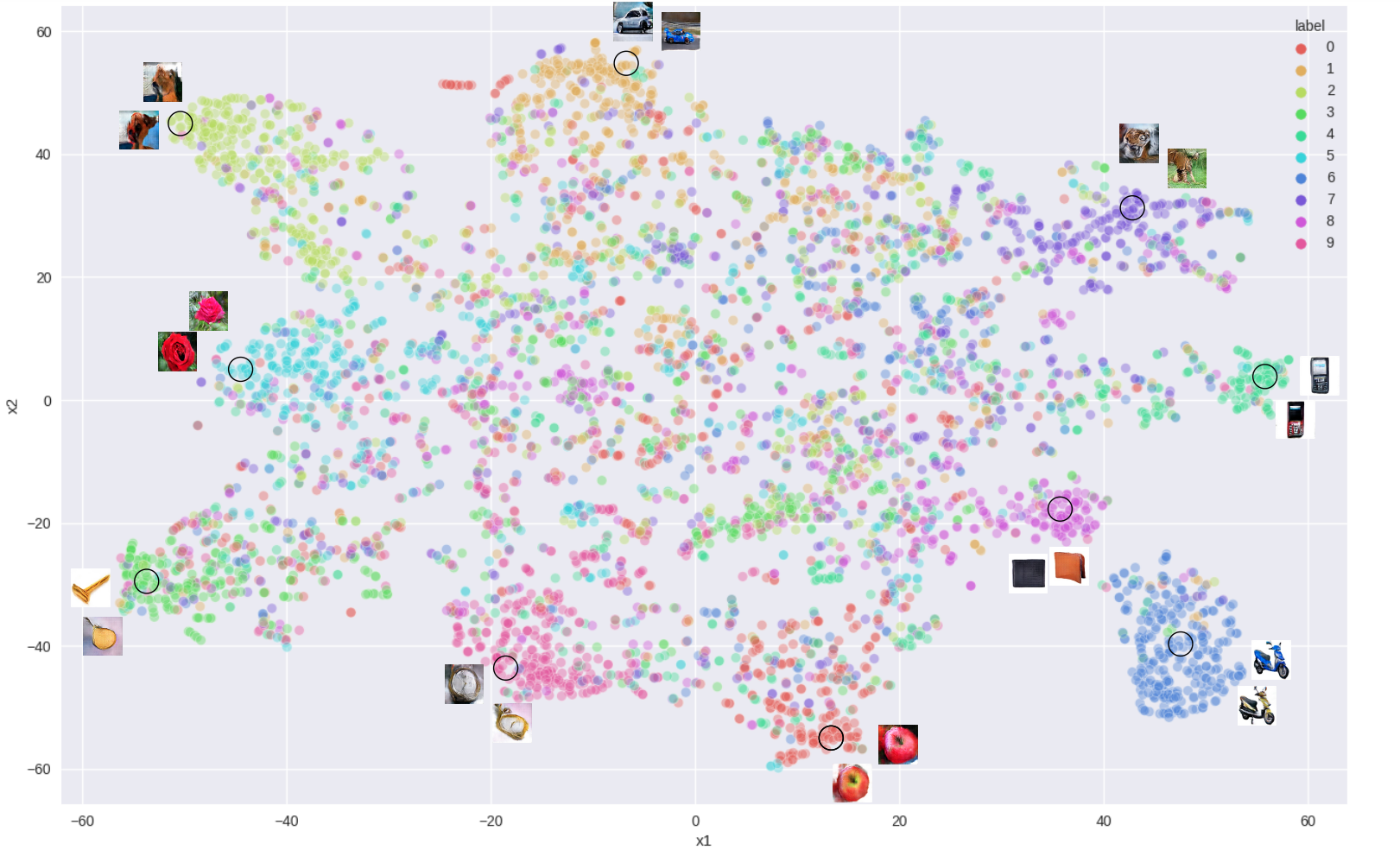}}
\end{minipage}
\caption{t-SNE \cite{van2008visualizing} visualization of Object test dataset \cite{Kumar2018} EEG feature space which is learned using triplet loss with test k-means accuracy 0.53. Each cluster's equivalent EEG-based generated images are also visualized in this plot.}
\label{fig:res_feature}
\vspace{-3mm}
\end{figure}

\begin{figure}[!t]

%
\begin{minipage}[b]{0.5\linewidth}
  \centering
  \centerline{\includegraphics[width=4.0cm, height=6.4cm]{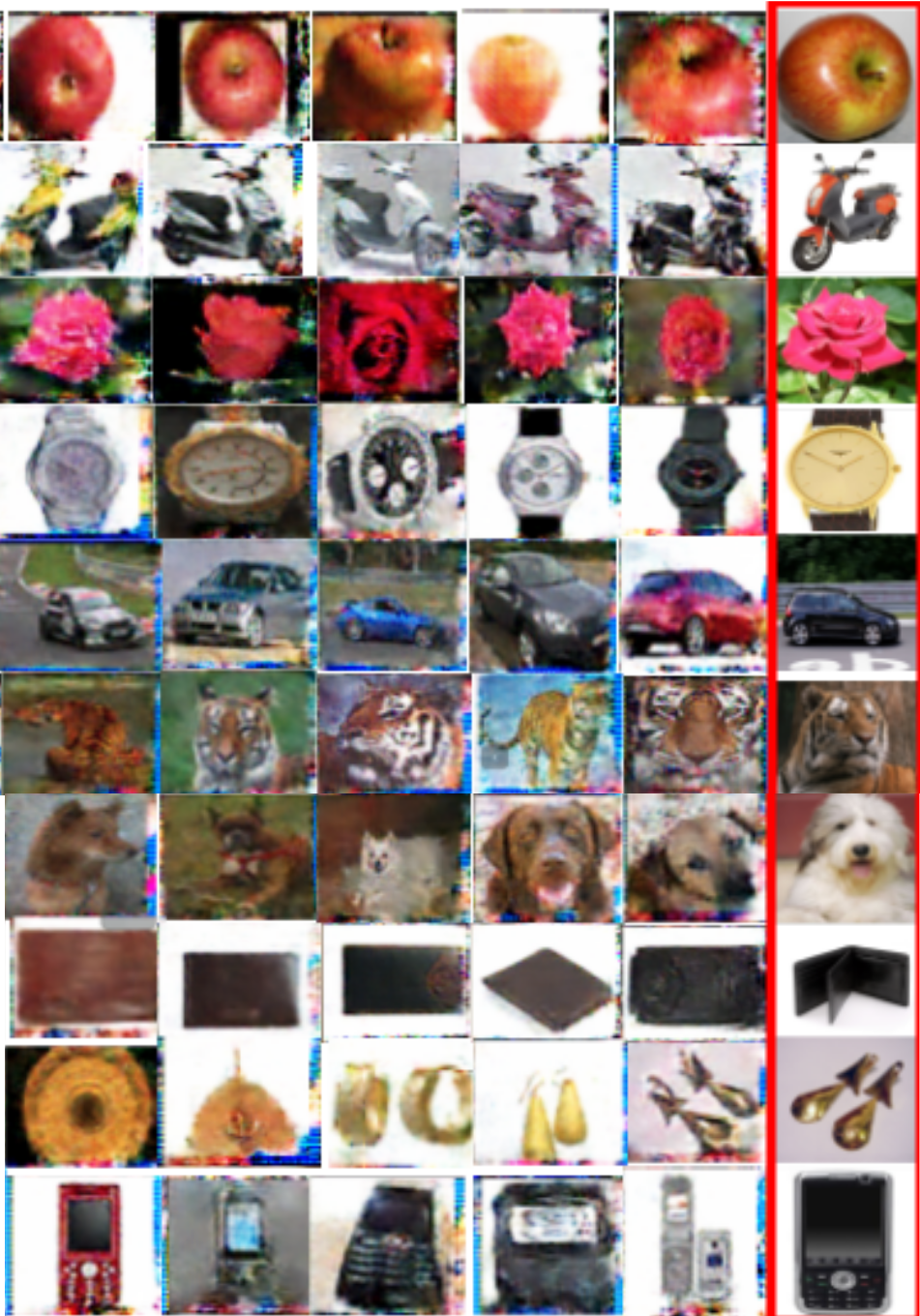}}
  \centerline{(a) ThoughtViz \cite{thoughtviz}}\medskip
\vspace{-3mm}
\end{minipage}
\begin{minipage}[b]{0.5\linewidth}
  \centering
  \centerline{\includegraphics[width=4.0cm, height=6.4cm]{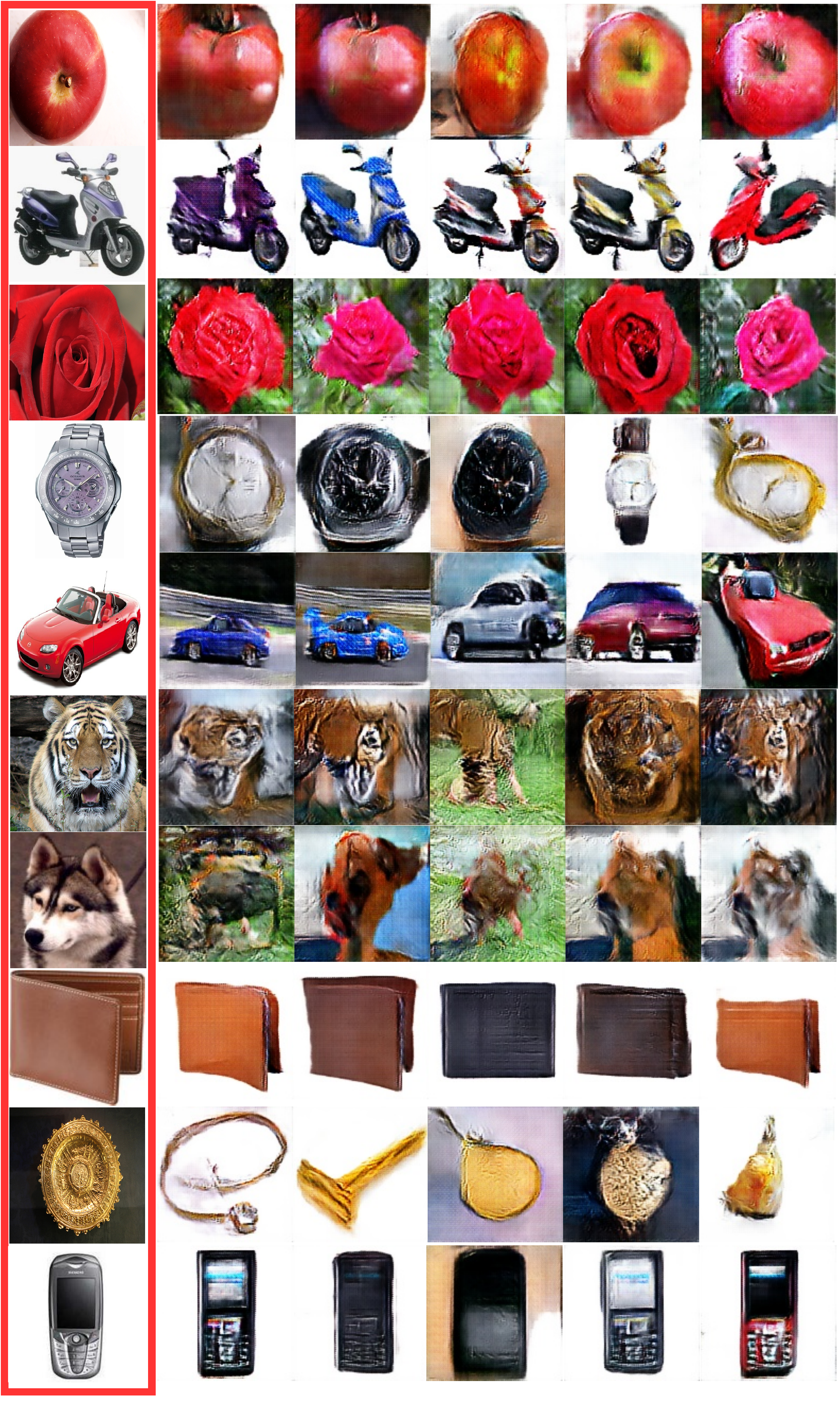}}
  \centerline{(b) EEG2Image (Ours)}\medskip
\vspace{-3mm}
\end{minipage}
\caption{Qualitative comparison between the images generated by EEG signals using the ThoughtViz method (left) and our proposed framework (right). Images in the red bounding box are the sample images from the Object test dataset \cite{Kumar2018}. These images are visualized by the participants, and the respective brain activity EEG signals is used here for reconstruction.}
\label{fig:res_imagenet10}
\end{figure}

\textbf{Feature Extraction.} Recent works \cite{palazzo2020decoding} have shown that the contrastive learning-based approach outperforms the supervised setting in the case of generalized feature learning for downstream tasks such as object detection, classification, saliency map from EEG signals, etc. Building on this, we have used a triplet loss-based contrastive learning \cite{schroff2015facenettriplet} approach in the proposed framework for EEG feature learning. Triplet loss aims to minimize the distance between the two data with the same labels and maximize the distance between the two data with different labels. To prevent the feature extraction network from squashing the representation of each data into a small cluster, a margin term is used in triplet loss. It ensures that the distance between the feature of the same label data is close to zero and greater than the margin for different label data. The formulation of triplet loss is as follows:
\begin{align}
    \min_{\theta}\mathbb{E}\big[ ||f_{\theta}(x^{a}) - f_{\theta}(x^{p})||_{2}^{2} - ||f_{\theta}(x^{a}) - f_{\theta}(x^{n})||_{2}^{2} + \beta \big]
    \label{eqn:1}
\end{align}
where, $f$ is parameterized function on $\theta$ that maps EEG signals to a feature space i.e. $f_{\theta}:\mathbb{R}^{C \times T} \xrightarrow{} \mathbb{R}^{128}$. The goal of Eqn. \ref{eqn:1} is to minimize the distance between anchor (a) EEG signal and positive (p) EEG signal of the same class as the anchor and maximize the distance between anchor EEG signal and negative (n) EEG signal of different class with margin distance. This formulation is also known as metric learning or contrastive learning. The idea behind using the formulation is to ensure that the EEG signals generated by the brain activity for similar images should be close to each other in the learned feature space \cite{tao2021clustering}. For learning better features, we have used semi-hard hard triplets, where the distance of the negative sample is more than positive but less than the margin, and also used an online hard-triplet mining strategy similar to \cite{schroff2015facenettriplet}.

\textbf{Image Generation.} In the proposed framework, we have used a Generative Adversarial Network (GAN) \cite{goodfellow2020generative} to synthesize the image from the extracted EEG feature. A GAN architecture consists of two sub-networks: Generator (G) and Discriminator (D). The purpose of a Generator is to learn the transformation between a latent distribution ($p_{\mathbb{Z}}$) and real-world data distribution ($p_{data}$). In our case, we assume latent distribution as an isotropic Gaussian $\mathcal{N}(0, I)$ from which we sample a noise vector $z \in \mathbb{R}^{128}$. The discriminator learns to distinguish real images from synthesized images. The complete GAN architecture is trained in a min-max optimization setting. Where the discriminator tries to maximize the score for real images $D(x)$ and minimize the score for generated images $D(G(z))$, in contrast to the discriminator, the generator tries to minimize $(1-D(G(z))$ and the minimizing of the term is only possible if generator synthesizes photorealistic images. The complete GAN optimization process can also be represented below:
\begin{align}
        \min_{G}\max_{D}V(D, G) = \mathbb{E}_{x \sim p_{data}(x)}[log(D(x))] + \notag\\
                                \mathbb{E}_{x \sim p_{\mathcal{Z}(z)}}[log(1-D(G(z))))]
\end{align}

\begin{table}[!t]
\center
\begin{tabular}{ll} \hline
\multicolumn{1}{l}{Method} & \multicolumn{1}{l}{Inception Score} \\ \hline
AC-GAN \cite{acgan}                    & 4.93                                \\ \hline
ThoughtViz \cite{thoughtviz}                & 5.43                                \\ \hline
EEG2Image (Ours)                       & \textbf{6.78}                               \\ \hline
\end{tabular}

\caption{Comparison of Inception Score values (on all classes of Object dataset \cite{Kumar2018}).}
\label{table:inceptioncomparision}
\vspace{-5mm}
\end{table}

\begin{figure}[!t]
%
\begin{minipage}[b]{0.5\linewidth}
  \centering
  \centerline{\includegraphics[width=3.7cm, height=4.2cm]{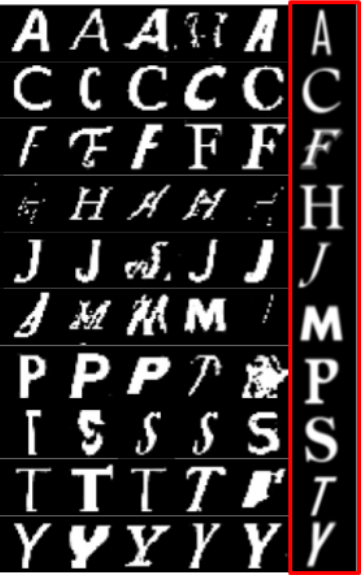}}
  \centerline{(a) ThoughtViz \cite{thoughtviz}}\medskip
\vspace{-3mm}
\end{minipage}
\hfill
\begin{minipage}[b]{0.5\linewidth}
  \centering
  \centerline{\includegraphics[width=3.7cm, height=4.2cm]{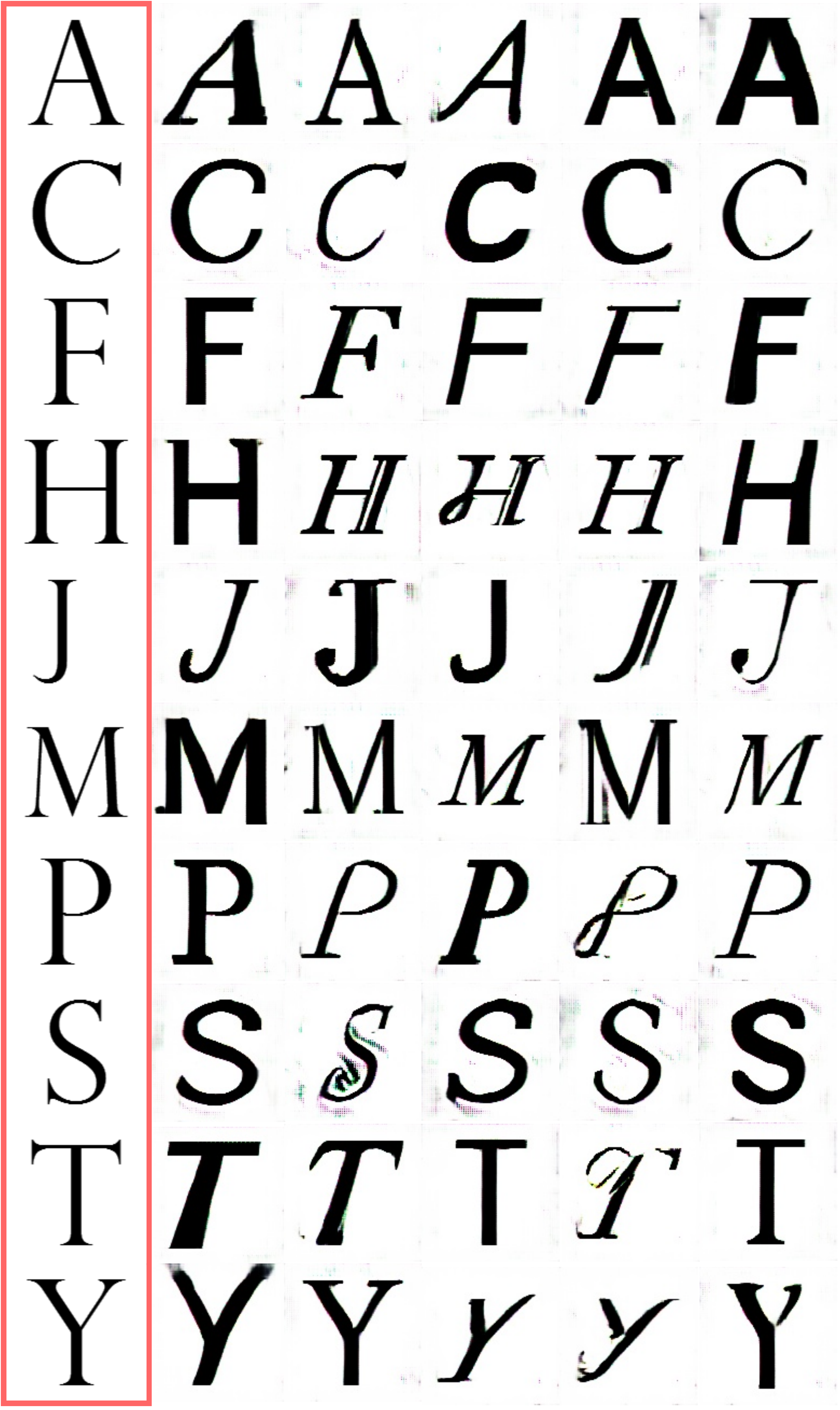}}
  \centerline{(b) EEG2Image (Ours)}\medskip
\vspace{-3mm}
\end{minipage}
\caption{Qualitative comparison between the images generated by the ThoughtViz method (left) and our proposed framework (right). Images in the red bounding box are the sample images from the Character test dataset \cite{Kumar2018}, these images are visualized by the participants, and the respective brain activity EEG signals is used here for reconstruction.}
\label{fig:res_alphabet}
\vspace{-3mm}
\end{figure}

\begin{figure*}[!t]

\begin{minipage}[b]{0.33\linewidth}
  \centering
  \centerline{\includegraphics[width=5.2cm]{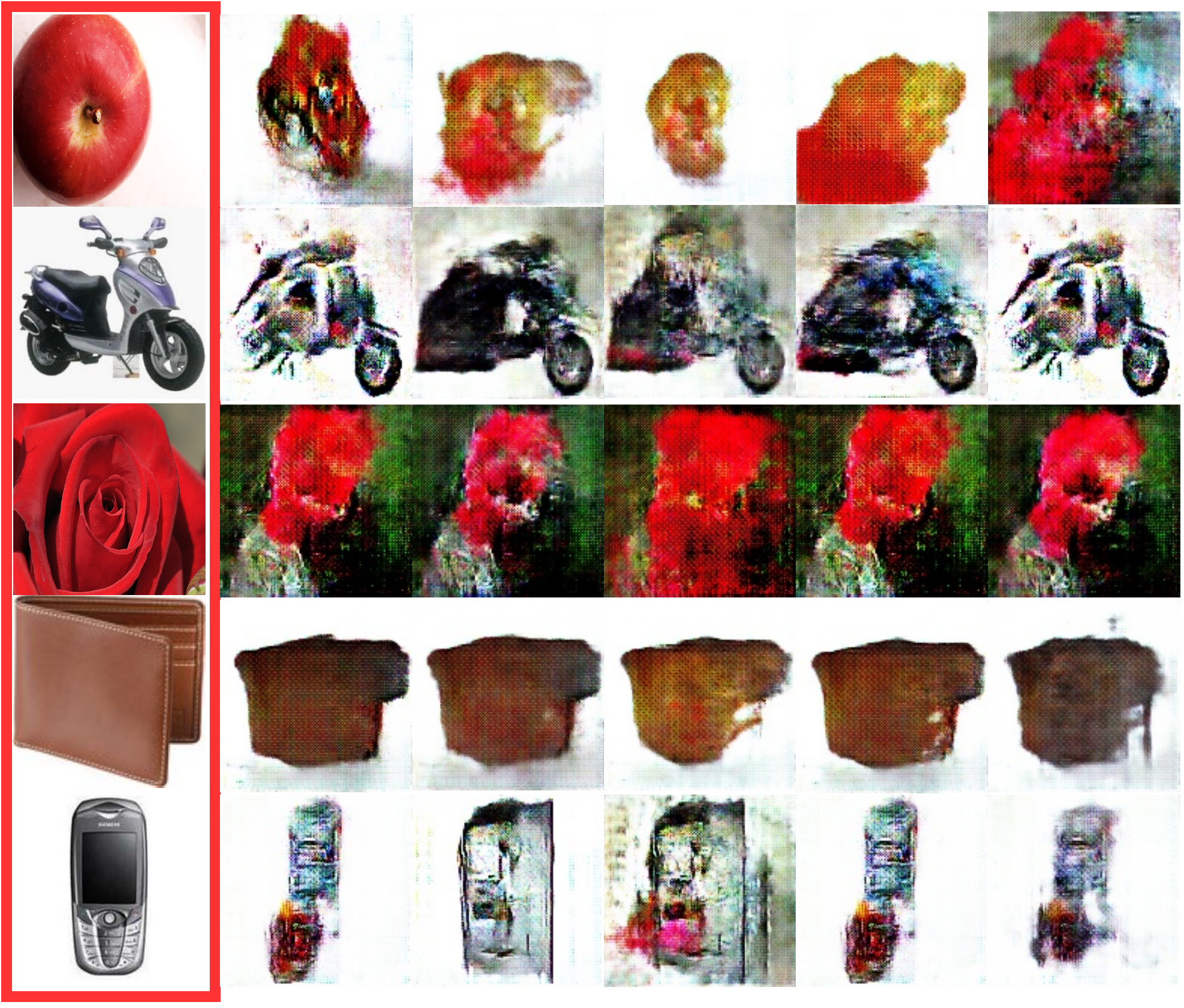}}
  \Centerline{(a) no \textcolor{orange}{modeloss} and \textcolor{orange}{dataaug}, \cpar inception score 3.61. }\medskip
  \vspace{-3mm}
\end{minipage}
\begin{minipage}[b]{.33\linewidth}
  \centering
  \centerline{\includegraphics[width=5.2cm]{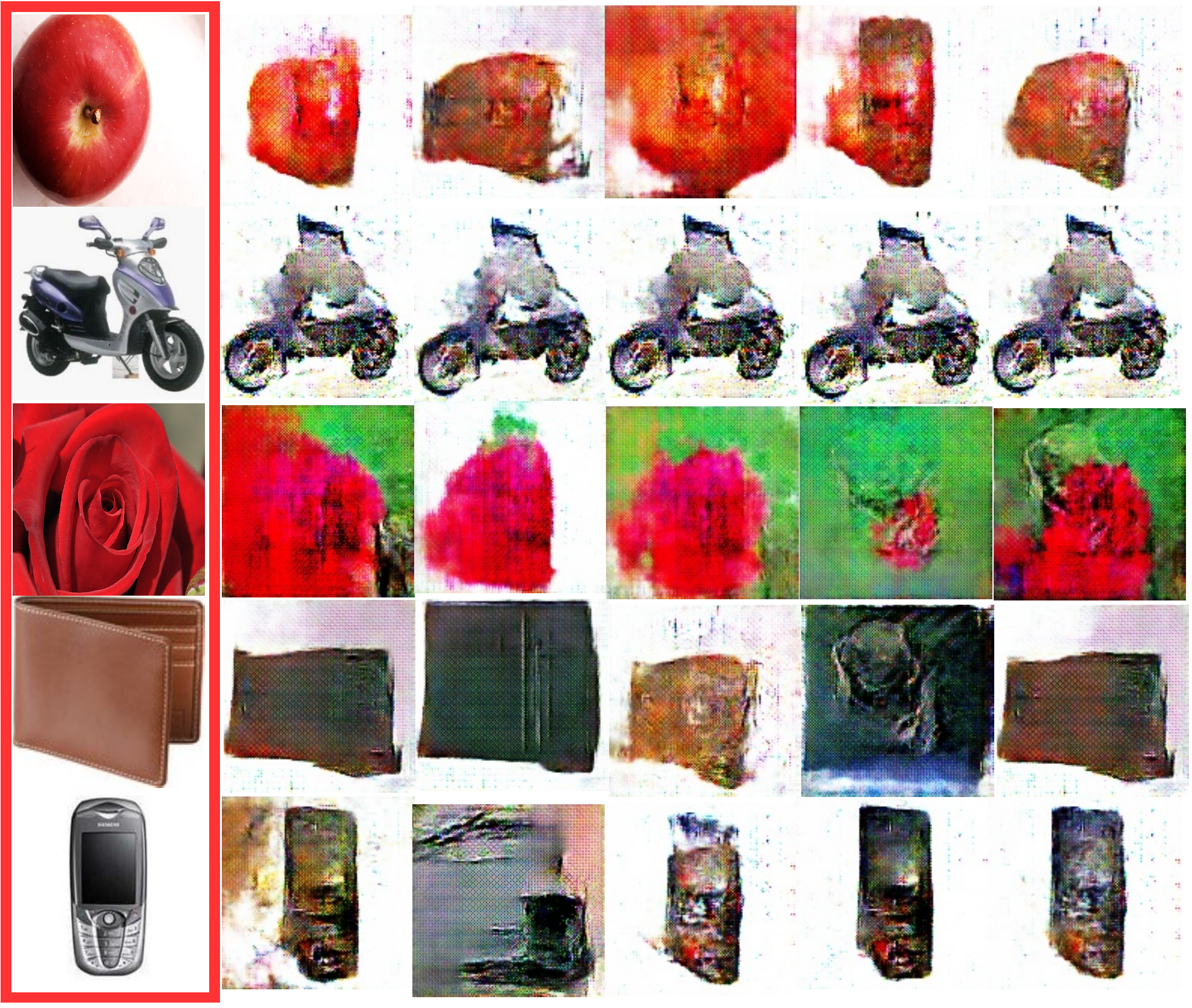}}
  \Centerline{(b) with \textcolor{green}{modeloss} and 
 no \textcolor{orange}{dataaug}, \cpar inception score 4.27. }\medskip
 \vspace{-3mm}
\end{minipage}
\begin{minipage}[b]{0.33\linewidth}
  \centering
  \centerline{\includegraphics[width=5.2cm]{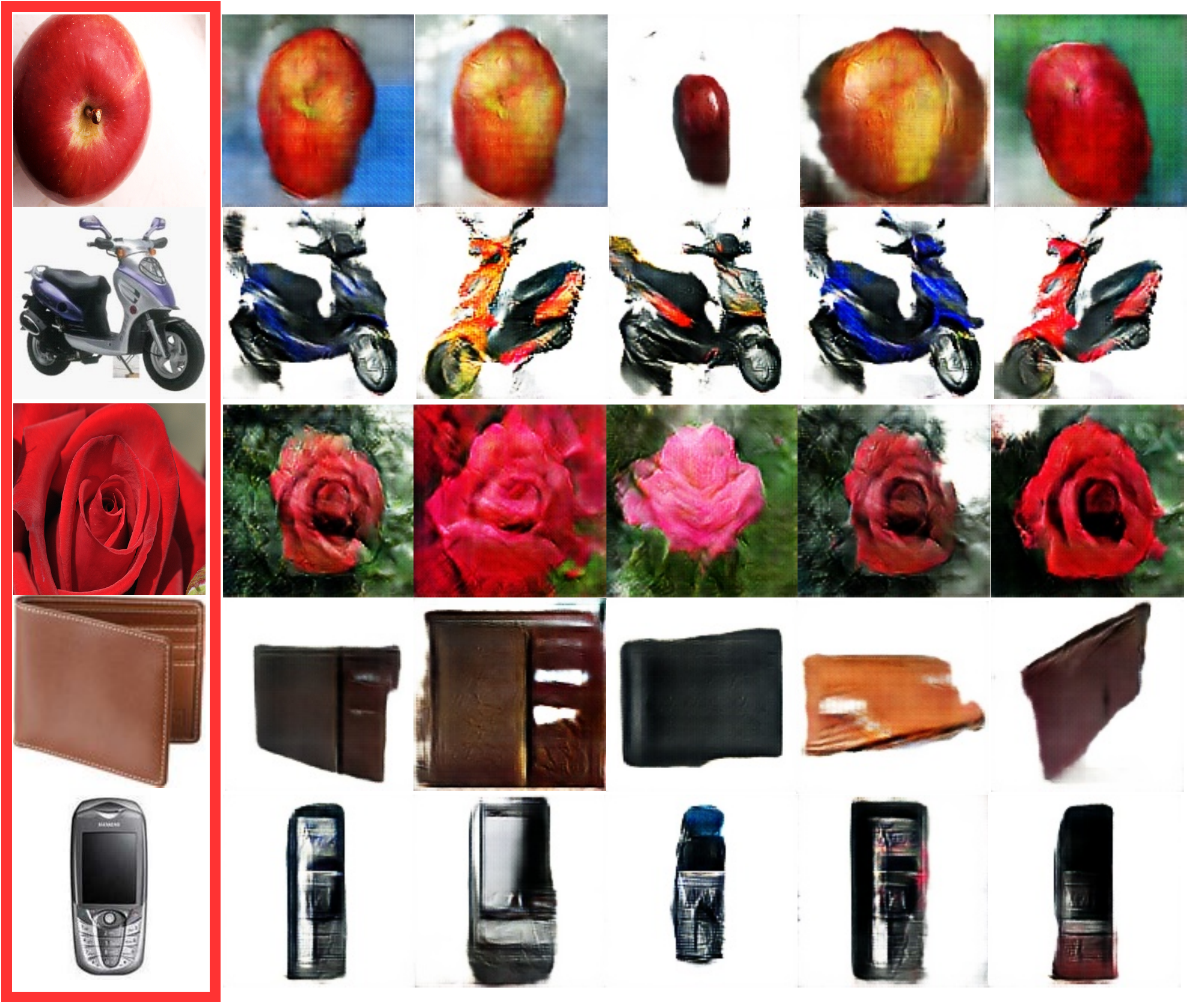}}
  \Centerline{(c) no \textcolor{orange}{modeloss} and with 
 \textcolor{green}{dataaug}, \cpar inception score 6.5. }\medskip
 \vspace{-3mm}
\end{minipage}
\caption{Ablation study showing the qualitative result on Object dataset \cite{Kumar2018} using different loss combinations for training the GAN network.}
\label{fig:ablation}
\end{figure*}
\begin{table*}[!t]
\resizebox{\textwidth}{!}{
\begin{tabular}{lllllllllll|l} \hline
Object Class                                                 & \begin{tabular}[c]{@{}l@{}}Apple\\ (n07739125)\end{tabular} & \begin{tabular}[c]{@{}l@{}}Car \\ (n02958343)\end{tabular} & \begin{tabular}[c]{@{}l@{}}Dog \\ (n02084071)\end{tabular} & \begin{tabular}[c]{@{}l@{}}Gold \\ (n03445326)\end{tabular} & \begin{tabular}[c]{@{}l@{}}Mobile \\ (n02992529)\end{tabular} & \begin{tabular}[c]{@{}l@{}}Rose \\ (n12620196)\end{tabular} & \begin{tabular}[c]{@{}l@{}}Scooter \\ (n03791053)\end{tabular} & \begin{tabular}[c]{@{}l@{}}Tiger \\ (n02129604)\end{tabular} & \begin{tabular}[c]{@{}l@{}}Wallet \\ (n04548362)\end{tabular} & \begin{tabular}[c]{@{}l@{}}Watch \\ (n04555897)\end{tabular} & All   \\ \hline
Mean                                                         & 6.09                                                        & 6.15                                                       & 6.99                                                       & 6.98                                                        & 7.33                                                          & 5.44                                                        & 5.81                                                           & 5.67                                                         & 6.48                                                          & 6.67                                                         & 6.78  \\
\begin{tabular}[c]{@{}l@{}}SD\end{tabular} & 0.05                                                        & 0.084                                                      & 0.031                                                      & 0.082                                                       & 0.030                                                         & 0.089                                                       & 0.077                                                          & 0.057                                                        & 0.086                                                         & 0.037                                                        & 0.086 \\ \hline
\end{tabular} 
}
\caption{Mean and standard deviation (SD) of Inception scores for each class of Objects dataset \cite{Kumar2018}.}
\label{table:perclassinception}
\vspace{-3.5mm}
\end{table*}
Similar to \cite{thoughtviz}, we aim to develop a framework that can utilize a small-size EEG dataset for generating images from EEG signals. To overcome the problem of small dataset \cite{thoughtviz} has used the trainable weighted Gaussian layer \cite{gurumurthy2017deligan}, which learns the mean ($\mu$) and variance ($\sigma$) for the encoded EEG signal. In this work, we follow a different strategy than \cite{thoughtviz}. Instead, we have used a Conditional DCGAN \cite{dcgan} architecture with the following modification 1) following the work of \cite{lim2017geometric}, we have used hinge loss for stable training of GAN, 2) we have added a differentiable data augmentation block between generator and discriminator which helps the network in learning from small datasize \cite{diffaug}, and 3) to ensure the mode diversity of GAN we have also used a mode seeking regularization as proposed in work \cite{mao2019mode}. In work \cite{lim2017geometric}, Lim \emph{et al.} have shown that the vanilla formulation of GAN suffers from mode collapse and unstable training problems. To solve these issues, they formulated an SVM separating hyperplane approach, which is known as GAN Hinge Loss as below:
\begin{align}
    \mathcal{L}_{D} &= \mathbb{E}_{x \sim p_{data}(x)}[max(0, 1 - D(x))] \text{ }+ \notag\\
                       & \qquad\qquad \mathbb{E}_{x \sim p_{\mathcal{Z}(z)}}[max(0, 1 + D(G(z)))] \\
    \mathcal{L}_{G} &= -\mathbb{E}_{x \sim p_{\mathcal{Z}(z)}}[D(G(z))]               
\end{align}

Training a GAN for synthesizing photorealistic images requires a large number of data \cite{diffaug}, and other deep learning approaches also face the same data scarcity issues. Zhao et al. \cite{diffaug} in their work have shown that the problem of sparse data for training a GAN can be resolved by adding a Differentiable Data Augmentation (DiffAug) block between the generator and discriminator, which is illustrated in Fig.\ref{fig:network}(b). The issue with sparse data is discriminator can easily memorize the data, which causes the vanishing gradient problem for the generator. The data augmentations we have used for our GAN network are translation and color jittering. The final loss term we aim to optimize for the proposed EEG2Image is given below:
\begin{align}
    \mathcal{L}_{D} &= \mathbb{E}_{(x,\psi) \sim p_{data}(x)}[max(0, 1 - D(\textcolor{red}{T}(x), \psi))] \text{ }+ \notag\\
                       & \qquad \mathbb{E}_{x \sim p_{\mathcal{Z}(z)}, \psi \sim p_{data}(x)}[max(0, 1 + D(\textcolor{red}{T}(G(z, \psi)), \psi))] \\
    \mathcal{L}_{ms} &= \min_{G} \bigg( \frac{d_{I}(G(\psi, z_{1}), G(\psi, z_{2}))}{d_{z}(z_{1}, z_{2})} \bigg)^{-1} \\
    \mathcal{L}_{G} &= -\mathbb{E}_{x \sim p_{\mathcal{Z}(z)}, \psi \sim p_{data}(x)}[D(\textcolor{red}{T}(G(z, \psi)), \psi)] + \alpha * \mathcal{L}_{ms}
\end{align}
where $\mathcal{L}_{D}$ is discriminator loss, $\mathcal{L}_{G}$ is generator loss, $\mathcal{L}_{ms}$ is mode seeking regularizer term \cite{mao2019mode},  $\textcolor{red}{T}$ is DiffAugment \cite{diffaug} function, $\psi$ is EEG feature vector and $\alpha$ is regularizer weight term which kept as $1.0$ for all the experiments.
\vspace{-0.2cm}
\section{Experiments and Results}
\label{sec:expnresult}
In the first part of this section, we will discuss the experimental setup we used to train the feature extraction and generative network, including the dataset. Later in this section, we will discuss all the ablation studies done to justify choices for the proposed framework.

\textbf{Datasets.} We have used the EEG data from \cite{Kumar2018}. This dataset consists of EEG signals for $3$ different subjects: Digits, Characters, and Objects. In our study, we have only used Characters and Object data because these are more diverse and complex data to show the effectiveness of the proposed framework. The Characters dataset consists of ten English alphabet classes and the subset of Chars74K \cite{de2009character}. Similarly, the Objects dataset consists of ten different object classes and the subset of ImageNet \cite{deng2009imagenet}. While collecting brain activity EEG signals of the participants, they were asked to think about one of these characters/objects at a time. To record the EEG signals, Emotiv EPOC+ \cite{emotivEPOCChannel} device is used, which has $14$ channels with a sampling rate of $128$ Hz per channel.  For each dataset, $23$ participants were asked to visualize every ten classes. Thus we have $230$ EEG samples per dataset. For our work, we have used the EEG data provided by the authors with train-test splits \cite{thoughtviz}. We would like to thank the authors for making it publicly available.

\textbf{EEG2Feature.} The first stage of our proposed framework is to convert EEG signals into useful features for image generation. For this, we design two regimes. In the first regime, we train a classification network for extracting EEG features as done in \cite{thoughtviz}. The classifier is a LSTM \cite{hochreiter1997long} network with $128$ hidden units using softmax cross-entropy loss. We use k-means clustering \cite{Jin2010kmean} as a metric for the learned EEG feature, i.e., higher k-means accuracy implies better learned representation \cite{tao2021clustering}. The first regime gives us $74.3\%$ \& $75.4\%$ classification accuracy on test data of Object dataset \& Character dataset \cite{Kumar2018} and k-means accuracy of $17.8\%$ and $16.3$\%, further we plot t-SNE map \cite{van2008visualizing} to visualize the clustering of test data features from Object dataset in Fig. \ref{fig:class_feature}. For the second regime, we used a contrastive learning approach to learning the feature of an EEG signal. As discussed in the Sec. \ref{sec:method} we used semi-hard triplet loss for training the LSTM \cite{hochreiter1997long} network with $128$ hidden units. The goal of triplet loss is to structured the feature space in such a way that positive pairs are in close proximity to each other while negative pairs are positioned far apart. The k-means accuracy we got on the test data of the Object dataset is $53\%$, and the Character dataset is $49\%$. Further, we plot t-SNE map \cite{van2008visualizing} to visualize the clustering of test data features from Object dataset Fig. \ref{fig:res_feature}. We can see that the k-means accuracy and t-SNE plot are better for the second regime. Therefore we decided to use the contrastive learning method as an EEG feature extractor for our proposed framework.

\textbf{Feature2Image.} The second stage of our proposed framework is to synthesize photorealistic images from extracted EEG features using the first stage. For synthesizing the image, we have used Conditional DCGAN \cite{dcgan} with modification as discussed in Sec. \ref{sec:method}. We have used Inception Score (IS) \cite{inceptionscore2016} as a metric for image quality comparison with other methods. Table \ref{table:inceptioncomparision} shows our proposed GAN method performed better in synthesizing the images from less number of EEG data. In Table \ref{table:perclassinception} we have shown per class inception score for test data of Object dataset \cite{Kumar2018}. We also performed the qualitative analysis of synthesized images for both the dataset Object and Character, which are shown in Figs. [\ref{fig:res_imagenet10}, \ref{fig:res_alphabet}]. We have performed several ablation studies to verify the importance of each loss in training the GAN network for the proposed framework. For that, we trained the proposed conditional GAN (cGAN) for three different regimes on the Object dataset \cite{Kumar2018}. In the first regime, we train the cGAN without mode seeking regularization and DiffAugment, shown in Fig. \ref{fig:ablation}(a). which has an inception score of $3.61$. In the second regime, we have added a mode seeking regularization term only and trained the cGAN from scratch, Fig. \ref{fig:ablation}(b) shows improvement with an inception score of $4.27$. In the third and last regime, we train the cGAN with the DiffAugment block, showing a large improvement in the synthesized image as shown in Fig. \ref{fig:ablation}(c) with an inception score of $6.5$. Based on these experiments, we used both the mode-seeking regularization term and the DiffAugment block in the proposed framework.


\vspace{-0.5cm}
\section{Conclusion}
\label{sec:conclusion}
\vspace{-0.2cm}
This work proposes a framework that uses a small-sized dataset for generating images from brain activity EEG signals. Our proposed framework has a better inception score than the previously proposed method for the small-sized EEG dataset and synthesized images of size $128 \times 128$. The framework consists of a contrastive learning approach to learn the good features of EEG data, which is empirically shown to perform better than the softmax-based supervised learning method. We have performed several ablation studies to demonstrate the effectiveness of modified GAN loss function in synthesizing high-quality images. As future work, we plan to tackle large-size EEG datasets and approach for complete self/un-supervised learning for extracting features from EEG data and image synthesis.

\vspace{-0.1cm}
\bibliographystyle{IEEEbib}
\small{
\bibliography{strings,refs}

\begin{thebibliography}{10}

\bibitem{miyawaki2008visual}
Yoichi Miyawaki, Hajime Uchida, Okito Yamashita, Masa-aki Sato, Yusuke Morito,
  Hiroki~C Tanabe, Norihiro Sadato, and Yukiyasu Kamitani,
\newblock ``Visual image reconstruction from human brain activity using a
  combination of multiscale local image decoders,''
\newblock {\em Neuron}, vol. 60, no. 5, pp. 915--929, 2008.

\bibitem{sakkalis2011applied}
Vangelis Sakkalis,
\newblock ``Applied strategies towards eeg/meg biomarker identification in
  clinical and cognitive research,''
\newblock {\em Biomarkers in medicine}, vol. 5, no. 1, pp. 93--105, 2011.

\bibitem{bamdad2015application}
Mahdi Bamdad, Homayoon Zarshenas, and Mohammad~A Auais,
\newblock ``Application of bci systems in neurorehabilitation: a scoping
  review,''
\newblock {\em Disability and Rehabilitation: Assistive Technology}, vol. 10,
  no. 5, pp. 355--364, 2015.

\bibitem{kavasidis2017brain2image}
Isaak Kavasidis, Simone Palazzo, Concetto Spampinato, Daniela Giordano, and
  Mubarak Shah,
\newblock ``Brain2image: Converting brain signals into images,''
\newblock in {\em 25th ACM MM}, 2017.

\bibitem{wang2022open}
Zhenhailong Wang and Heng Ji,
\newblock ``Open vocabulary electroencephalography-to-text decoding and
  zero-shot sentiment classification,''
\newblock in {\em AAAI}, 2022, vol.~36, pp. 5350--5358.

\bibitem{carlson2013representational}
Thomas Carlson, David~A Tovar, Arjen Alink, and Nikolaus Kriegeskorte,
\newblock ``Representational dynamics of object vision: the first 1000 ms,''
\newblock {\em Journal of vision}, vol. 13, no. 10, pp. 1--1, 2013.

\bibitem{carlson2011high}
Thomas~A Carlson, Hinze Hogendoorn, Ryota Kanai, Juraj Mesik, and Jeremy
  Turret,
\newblock ``High temporal resolution decoding of object position and
  category,''
\newblock {\em Journal of vision}, vol. 11, no. 10, pp. 9--9, 2011.

\bibitem{das2010predicting}
Koel Das, Barry Giesbrecht, and Miguel~P Eckstein,
\newblock ``Predicting variations of perceptual performance across individuals
  from neural activity using pattern classifiers,''
\newblock {\em Neuroimage}, vol. 51, no. 4, pp. 1425--1437, 2010.

\bibitem{spampinato2017deep}
Concetto Spampinato, Simone Palazzo, Isaak Kavasidis, Daniela Giordano, Nasim
  Souly, and Mubarak Shah,
\newblock ``Deep learning human mind for automated visual classification,''
\newblock in {\em IEEE CVPR}, 2017, pp. 6809--6817.

\bibitem{schroff2015facenettriplet}
Florian Schroff, Dmitry Kalenichenko, and James Philbin,
\newblock ``Facenet: A unified embedding for face recognition and clustering,''
\newblock in {\em IEEE CVPR}, 2015, pp. 815--823.

\bibitem{mao2019mode}
Qi~Mao, Hsin-Ying Lee, Hung-Yu Tseng, Siwei Ma, and Ming-Hsuan Yang,
\newblock ``Mode seeking generative adversarial networks for diverse image
  synthesis,''
\newblock in {\em IEEE/CVF CVPR}, 2019.

\bibitem{diffaug}
Shengyu Zhao, Zhijian Liu, Ji~Lin, Jun-Yan Zhu, and Song Han,
\newblock ``Differentiable augmentation for data-efficient gan training,''
\newblock {\em NeurIPS}, vol. 33, pp. 7559--7570, 2020.

\bibitem{thoughtviz}
Praveen Tirupattur, Yogesh~Singh Rawat, Concetto Spampinato, and Mubarak Shah,
\newblock ``Thoughtviz: Visualizing human thoughts using generative adversarial
  network,''
\newblock in {\em Proceedings of the 26th ACM International Conference on
  Multimedia}, New York, NY, USA, 2018, MM '18, p. 950–958, ACM.

\bibitem{zheng2020decoding}
Xiao Zheng, Wanzhong Chen, Mingyang Li, Tao Zhang, Yang You, and Yun Jiang,
\newblock ``Decoding human brain activity with deep learning,''
\newblock {\em Biomedical Signal Processing and Control}, vol. 56, pp. 101730,
  2020.

\bibitem{zhang2019multi}
Xiang Zhang, Xiaocong Chen, Manqing Dong, Huan Liu, Chang Ge, and Lina Yao,
\newblock ``Multi-task generative adversarial learning on geometrical shape
  reconstruction from eeg brain signals,''
\newblock {\em arXiv preprint arXiv:1907.13351}, 2019.

\bibitem{fares2020brain}
Ahmed Fares, Sheng-hua Zhong, and Jianmin Jiang,
\newblock ``Brain-media: A dual conditioned and lateralization supported gan
  (dcls-gan) towards visualization of image-evoked brain activities,''
\newblock in {\em 28th ACM MM}, 2020, pp. 1764--1772.

\bibitem{palazzo2020decoding}
Simone Palazzo, Concetto Spampinato, Isaak Kavasidis, Daniela Giordano, Joseph
  Schmidt, and Mubarak Shah,
\newblock ``Decoding brain representations by multimodal learning of neural
  activity and visual features,''
\newblock {\em IEEE PAMI}, vol. 43, no. 11, pp. 3833--3849, 2020.

\bibitem{khare2022neurovision}
Sanchita Khare, Rajiv~Nayan Choubey, Loveleen Amar, and Venkanna Udutalapalli,
\newblock ``Neurovision: perceived image regeneration using cprogan,''
\newblock {\em Neural Computing and Applications}, vol. 34, no. 8, pp.
  5979--5991, 2022.

\bibitem{ye2022see}
Zesheng Ye, Lina Yao, Yu~Zhang, and Silvia Gustin,
\newblock ``See what you see: Self-supervised cross-modal retrieval of visual
  stimuli from brain activity,''
\newblock {\em arXiv preprint arXiv:2208.03666}, 2022.

\bibitem{van2008visualizing}
Laurens Van~der Maaten and Geoffrey Hinton,
\newblock ``Visualizing data using t-sne.,''
\newblock {\em JMLR}, vol. 9, no. 11, 2008.

\bibitem{Kumar2018}
Pradeep Kumar, Rajkumar Saini, Partha~Pratim Roy, Pawan~Kumar Sahu, and
  Debi~Prosad Dogra,
\newblock ``Envisioned speech recognition using eeg sensors,''
\newblock {\em Personal and Ubiquitous Computing}, vol. 22, no. 1, pp.
  185--199, Feb 2018.

\bibitem{tao2021clustering}
Yaling Tao, Kentaro Takagi, and Kouta Nakata,
\newblock ``Clustering-friendly representation learning via instance
  discrimination and feature decorrelation,''
\newblock {\em preprint arXiv:2106.00131}, 2021.

\bibitem{goodfellow2020generative}
Ian Goodfellow, Jean Pouget-Abadie, Mehdi Mirza, Bing Xu, David Warde-Farley,
  Sherjil Ozair, Aaron Courville, and Yoshua Bengio,
\newblock ``Generative adversarial networks,''
\newblock {\em Communications of the ACM}, vol. 63, no. 11, pp. 139--144, 2020.

\bibitem{acgan}
Augustus Odena, Christopher Olah, and Jonathon Shlens,
\newblock ``Conditional image synthesis with auxiliary classifier gans,''
\newblock in {\em ICML}. PMLR, 2017, pp. 2642--2651.

\bibitem{gurumurthy2017deligan}
Swaminathan Gurumurthy, Ravi Kiran~Sarvadevabhatla, and R~Venkatesh~Babu,
\newblock ``Deligan: Generative adversarial networks for diverse and limited
  data,''
\newblock in {\em IEEE CVPR}, 2017.

\bibitem{dcgan}
Alec Radford, Luke Metz, and Soumith Chintala,
\newblock ``Unsupervised representation learning with deep convolutional
  generative adversarial networks,''
\newblock {\em preprint arXiv:1511.06434}, 2015.

\bibitem{lim2017geometric}
Jae~Hyun Lim and Jong~Chul Ye,
\newblock ``Geometric gan,''
\newblock {\em arXiv preprint arXiv:1705.02894}, 2017.

\bibitem{de2009character}
Te{\'o}filo~Em{\'\i}dio De~Campos, Bodla~Rakesh Babu, Manik Varma, et~al.,
\newblock ``Character recognition in natural images.,''
\newblock {\em VISAPP (2)}, vol. 7, no. 2, 2009.

\bibitem{deng2009imagenet}
Jia Deng, Wei Dong, Richard Socher, Li-Jia Li, Kai Li, and Li~Fei-Fei,
\newblock ``Imagenet: A large-scale hierarchical image database,''
\newblock in {\em IEEE CVPR}. Ieee, 2009, pp. 248--255.

\bibitem{emotivEPOCChannel}
``{E}{P}{O}{C}+ - 14 {C}hannel {E}{E}{G} --- emotiv.com,''
  \url{https://www.emotiv.com/epoc/},
\newblock [Accessed 14-Oct-2022].

\bibitem{hochreiter1997long}
Sepp Hochreiter and J{\"u}rgen Schmidhuber,
\newblock ``Long short-term memory,''
\newblock {\em Neural computation}, vol. 9, no. 8, 1997.

\bibitem{Jin2010kmean}
Xin Jin and Jiawei Han,
\newblock {\em K-Means Clustering}, pp. 563--564,
\newblock Springer US, Boston, MA, 2010.

\bibitem{inceptionscore2016}
Tim Salimans, Ian Goodfellow, Wojciech Zaremba, Vicki Cheung, Alec Radford, and
  Xi~Chen,
\newblock ``Improved techniques for training gans,''
\newblock in {\em 30th NeurIPS}, Red Hook, NY, USA, 2016, NIPS'16, p.
  2234–2242, Curran Associates Inc.

\end{thebibliography}
}

\end{document}